\documentclass[journal=jacsat,manuscript=article]{achemso}
\usepackage[version=3]{mhchem} 
\usepackage[T1]{fontenc}       
\usepackage{graphicx}
\usepackage{epsfig}
\usepackage[autostyle]{csquotes}
\usepackage{achemso}
\setkeys{acs}{usetitle = true}

\author{J. L. Carini}
\affiliation{Department of Physics, University of Connecticut, Storrs, Connecticut 06269, USA}

\author{S. Kallush}
\affiliation{Department of Physics and Optical Engineering, ORT Braude, P.O. Box 78, Karmiel, Israel}
\alsoaffiliation{Department of Physical Chemistry and the Fritz Haber Research Center for Molecular Dynamics, The Hebrew University, 91094, Jerusalem, Israel}

\author{R. Kosloff}
\affiliation{Department of Physical Chemistry and the Fritz Haber Research Center for Molecular Dynamics, The Hebrew University, 91094, Jerusalem, Israel}

\author{P. L. Gould}
\affiliation{Department of Physics, University of Connecticut, Storrs, Connecticut 06269, USA}
\email{* phillip.gould@uconn.edu}
\affiliation{Department of Physics, University of Connecticut, Storrs, Connecticut 06269, USA}

\title{Efficient Formation of Ultracold Molecules with Chirped Nanosecond Pulses}

\abbreviations{abbreviations? print here}

\begin{document}



\begin{abstract}
  We describe experiments and associated quantum simulations involving the production of ultracold $^{87}$Rb$_{2}$ molecules with nanosecond pulses of frequency-chirped light. With appropriate chirp parameters, the formation is dominated by coherent processes. For a positive chirp, excited molecules are produced by photoassociation early in the chirp, then transferred into high vibrational levels of the lowest triplet state by stimulated emission later in the chirp.  Generally good agreement is seen between the data and the simulations. Shaping of the chirp can lead to a significant enhancement of the formation rate. Further improvements using higher intensities and different intermediate states are predicted. 
	\end{abstract}
\section{1. Introduction}

Ultracold molecules \cite{Krems09,Jin12,Carr15} have become a topic of significant interest in recent years, with much effort aimed at their production and manipulation. This interest is due in part to their potential use in both basic science, such as novel dipolar systems, quantum degenerate gases, and tests of fundamental symmetries, as well as in the more applied realm, such as ultracold chemistry, precision spectroscopy and quantum computing. Because they have multiple degrees of freedom, including electronic state, vibration, rotation, electron spin, and nuclear spin, molecules are more difficult to cool and manipulate than atoms. The laser cooling techniques \cite{Metcalf99} so successfully used with atoms are not easily extended to molecules. Nevertheless, two general production methods for cold molecules have emerged: \enquote{direct}, such as buffer gas cooling, electrostatic slowing, and laser cooling; and \enquote{indirect}, such as photoassociation and magnetoassociation, where the molecules are assembled from already ultracold atoms.

	We are particularly interested in photoassociation (PA) \cite{Thorsheim87,Stwalley99,Jones06,Stwalley09}, where laser light causes a free-bound transition from the low-energy continuum to a bound excited state of the molecule. Subsequent transfer to lower-lying states, notably the electronic ground state, usually occurs by spontaneous emission. This process is not only incoherent, but also populates a broad range of vibrational levels, and in some cases, the continuum. An important question is whether the techniques of quantum control can be employed to improve upon this scheme \cite{Koch12,Vala00,Luc-Koenig04a,Luc-Koenig04b,Koch06a,Poschinger06,Koch06b,Brown06a,Koch06c,Mur-Petit07,Kallush07a,Kallush08,Koch08,Koch09,Kallush07b,Huang14,Tomza12,Salzmann06,Brown06,Salzmann08,McCabe09}, thus enabling coherent and efficient ultracold molecule formation. We show that nanosecond pulses of frequency-chirped light are a promising step in this direction.
	
	Quantum control \cite{Rice92,Brumer92,Brif10,Rice00,Shapiro03} is an area of interest in its own right, with applications ranging from laser-controlled chemistry to quantum information. It is based on the use of interfering pathways to enhance an objective, with ultrafast lasers as the usual tool. In our case, the objective is ultracold molecule formation, and because we are dealing with very low temperatures, the time scale is slowed from femtosecond/picoseconds to nanoseconds. We note that, in complimentary work, photoassociative molecular formation at higher temperatures and with ultrafast pulses has also been  investigated \cite{Levin15}. 
	
	We recently reported on the enhancement of ultracold molecule formation using nanosecond pulses of light with a shaped frequency chirp \cite{Carini15b}. We found that a positive piecewise linear chirp, where the frequency evolution is slow, then fast, then slow again, outperformed a positive linear chirp, a negative linear chirp, and unchirped light. In the present work, we provide a more comprehensive account of this earlier report, providing more details on both the experiment and the accompanying quantum simulations. We also discuss possible future improvements using higher intensities and different excited states.
	
	The paper is organized as follows. In Section 2, we discuss the quantum simulations, including the relevant molecular potentials, the Hamiltonian and the numerical methods for its solution, and the procedures needed to compare the theoretical results with the data. In Section 3, we describe the various features of the experiment and present the experimental data. Section 4 compares the theoretical and experimental results and discusses the interpretation which follows from this comparison. Future prospects are presented in Section 5, and Section 6 comprises concluding remarks.

\section{2. Theory}

In order to interpret experimental data, as well as to provide guidance to the experiments, we carry out quantum simulations of the ultracold collisional dynamics. Some aspects are described in earlier publications \cite{Carini13,Carini15,Carini15b}; here we give a more complete account.

\begin{figure}
    \centering 
    \includegraphics[width=8.25cm]{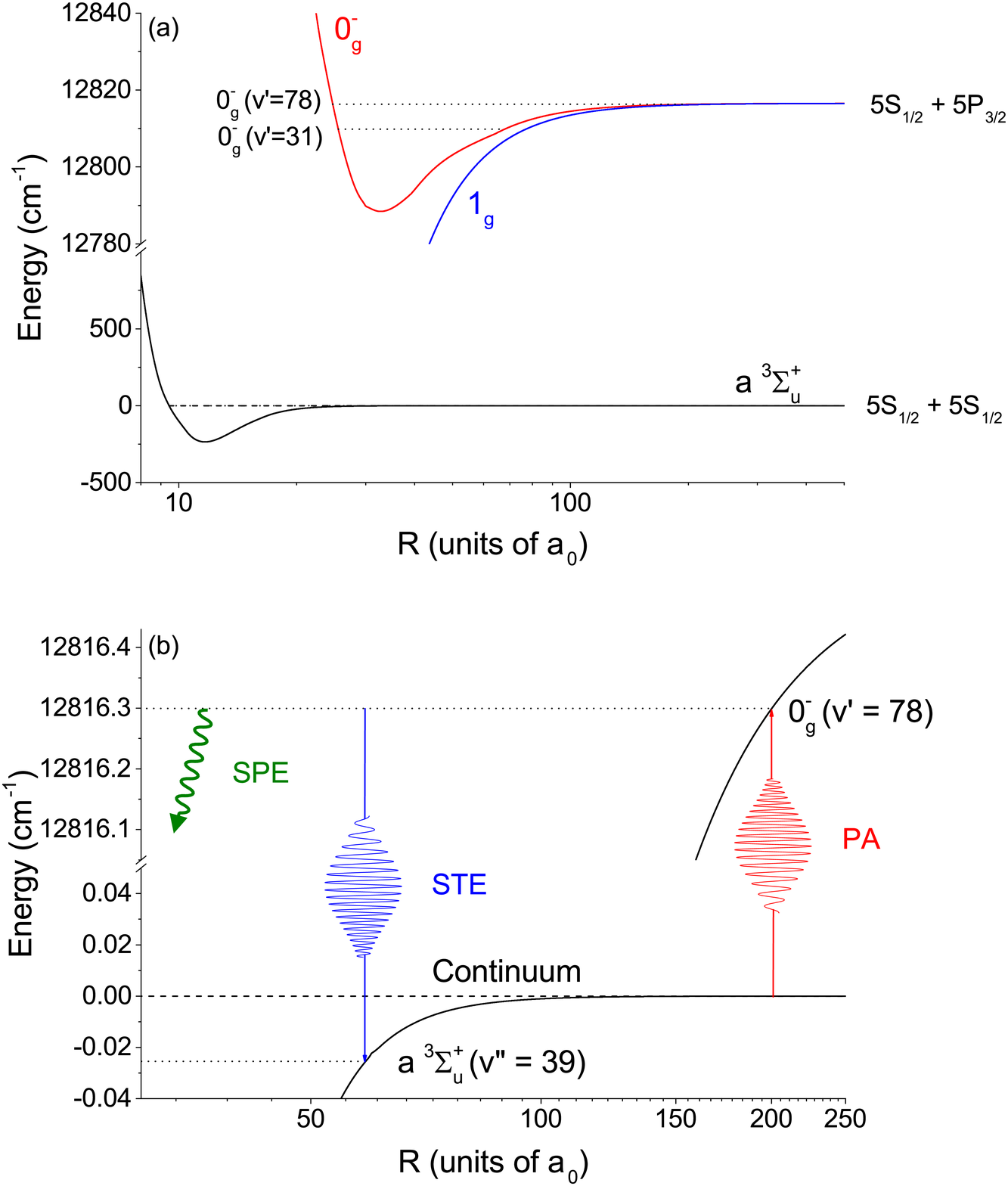} 
    \caption{(a) Molecular potentials relevant to the chirped molecule formation. Note that the horizontal axis is logarithmic and that there is a break in the vertical axis.  (b) Expanded view of the important states near dissociation: 0$_{g}^{-}$ (v'=78) bound by 7.79 GHz; and \textit{a} $^{3}\Sigma_{u}^{+}$ (v''=39) bound by 764 MHz. Also shown are the transitions driven by the chirped light: photoassociation (PA); and stimulated emission (STE). Spontaneous emission (SPE) of the excited state can also occur.}
\end{figure}

	Based on the photoassociation line chosen in the experiment, which has been assigned to the 0$_{g}^{-}$ state \cite{Kemmann04}, we include the following molecular potentials, denoted by j:  \textit{a} $^{3}\Sigma_{u}^{+}$ (j=g), the lowest triplet (metastable) state, often referred to as the \enquote{ground} state; 0$_{g}^{-}$ (j=0), the pure-long-range excited state below the 5s$_{1/2}$+5p$_{3/2}$ asymptote; and 1$_{g}$ (j=1), another excited state below this asymptote. This set of states, as shown in Fig. 1, is based on the g-u selection rules and the ranges of energies and internuclear separations, R, which are probed. In this picture, since the $^{87}$Rb atoms in the magneto-optical trap (MOT) are primarily in the F=2 hyperfine level, the \textit{a} $^{3}\Sigma_{u}^{+}$ potential is assumed to correlate to the (F=2) + (F=2) asymptote. Compared to 0$_{g}^{-}$, the 1$_{g}$ state has similar strength for the PA transition, but much weaker coupling to the \textit{a} $^{3}\Sigma_{u}^{+}$ (v''=39) target state.  Although it contributes minimally to the  target state population, it is included for completeness. We use the potentials for these states calculated in reference \cite{Gutterres02}. The \textit{a} $^{3}\Sigma_{u}^{+}$ state is adjusted slightly to give the correct s-wave scattering length \cite{Roberts98,Geltman01}, while the excited-state asymptotes are adjusted to give the measured vibrational spacing \cite{Kemmann04}.
	
	We solve the time-dependent Schr$\ddot{o}$dinger equation with a dressed-state Hamiltonian which includes the kinetic energy operator $\hat{T}$ and the relevant molecular potentials V$_{j}$ (j=g,0,1) discussed above. Rotational barriers are added to V$_{g}$ for partial waves up to J=5.
	
	The laser-induced coupling between the \textit{a} $^{3}\Sigma_{u}^{+}$ (g) and j=0,1 excited states is time dependent due to the variation of both the frequency (chirp) and the amplitude (pulse envelope):

\begin{equation}
 \hbar  \Omega_{j}(t) = \mu_{gj}\epsilon_{0}e^{[-\frac{(t-t_{center})^{2}}{2\sigma^{2}}+i\widetilde{\omega}(t)(t-t_{center})]}
\end{equation} Here, $\mu_{gj}$ are the transition dipole moments, taken to be independent of the internuclear separation R, $\epsilon_{0}$ is the peak electric field which occurs at time t = t$_{center}$, $\sigma$ = 6.4 ns for our 15 ns FWHM intensity pulse, and $\widetilde{\omega}$(t) are instantaneous frequency offsets from the main PA resonance (0$_{g}^{-}$ (v'=78)), taken from the measured frequency chirps.

Because our nanosecond time scales are quite long compared to those of the internal dynamics, we transform to a restricted basis set of vibrational levels in which the Hamiltonian can be written:

\begin{equation}
\hat{H}=
\begin{pmatrix}
\hat{H}_{g} & \hbar \hat{\Omega}_{0}(t) &  \hbar \hat{\Omega}_{1}(t) \\ \hbar \hat{\Omega}_{0}^{*}(t) & \hat{H}_{0} & 0 \\ \hbar \hat{\Omega}_{1}^{*}(t) & 0 & \hat{H}_{1}
\end{pmatrix}
\end{equation} Here the $\hat{H}_{j}$ are the vibrational energies of the various molecular states, calculated by the mapped Fourier grid method \cite{Kallush06}, and the couplings between vibrational levels, $\hat{\Omega}_{j}$, include the Franck-Condon factors (FCFs). Because we are dealing with small chirp ranges, and are quite close to the excited-state asymptotes, the excited basis sets are limited to a bandwidth of $\sim$15 GHz, which corresponds to 21 levels for $0_{g}^{-}$ and 23 levels for $1_{g}$. The \textit{a} $^{3}\Sigma_{u}^{+}$ basis set includes the 11 highest bound levels, spanning 278 GHz, and a narrow continuum of 16 MHz corresponding to an energy spread of E/k$_{B}$ = 0.77 mK, where k$_{B}$ is Boltzmann's constant. Even with these restricted basis sets, the calculations are computationally heavy since the dynamics must be followed for >100 ns, compared to the <100 ps durations for typical coherent control calculations. As in our earlier work \cite{Carini13,Carini15,Carini15b}, we exclude the contributions of the barely-bound (39 MHz) \textit{a} $^{3}\Sigma_{u}^{+}$ (v''=40) level, which is easily photodestroyed and therefore likely not detected. 

	Although we are operating on faster time scales than in our initial molecular work \cite{Carini13}, the effects of spontaneous emission (SPE) cannot be completely ignored. Our pulses are 15 ns FWHM, which is somewhat shorter than the 26.2 ns and 22.8 ns excited-state lifetimes for the $0_{g}^{-}$ and $1_{g}$ states, respectively \cite{Julienne91,Gutterres02}. We add multiple sink channels to account for decay into the various bound states and the continuum of \textit{a} $^{3}\Sigma_{u}^{+}$. These decay products are included in final populations, but not allowed to participate in subsequent dynamics. The limitations of this model are not important to the states of interest, except possibly for unchirped pulses, where spontaneous emission dominates. For the $0_{g}^{-}$ (v'=78) excited state, discussed in Sects. 3 and 4, >93$\%$ of the SPE to bound states ends up in v''=37-39, which are separated in energy by less then the 0.2 cm$^{-1}$ detection laser linewidth.  For the $0_{g}^{-}$ (v'=31) excited state, discussed in Sect. 5, this fraction is 43$\%$.  In this case, the calculated SPE populations are likely an overestimate of what would be seen in the experiment. In both cases, we find that the  \textit{a} $^{3}\Sigma_{u}^{+}$ population for positive chirps to be dominated by STE to v''=39, rendering the details of the SPE less important.

	The initial state is taken to be a box-normalized scattering state at energy k$_{B}$T with T=150 $\mu$K. At each time step in the calculation, the populations in the various states are generated, allowing the dynamics to be followed. For given chirp parameters and a fixed peak intensity, the final (200 ns after the beginning of the chirp)  \textit{a} $^{3}\Sigma_{u}^{+}$state probabilities must be converted to the time-averaged formation rate, R, the quantity measured in the experiment. This conversion requires several steps. First, we account for the thermal ensemble, as described in reference \cite{Koch06a}.  We find the number of molecules per pulse:\begin{equation}
N_{mol,J} = \frac{\pi^{2} \hbar^{3} Nn P_{E_{0},J}}{\mu^{3/2} \sqrt{E_{0}} \left. \frac{dE}{dn}\right|_{E_{0}}} \frac{}{},
\end{equation} where n is the atomic density, N is the atom number, and $\left. \frac{dE}{dn}\right|_{E_{0}}$ is the density of energy states evaluated at E$_{0}$. Then we multiply by the chirp repetition rate. Next, we spatially average over the density distribution of the trapped atoms. This distribution is measured to be Gaussian with 1/e$^{2}$ radii of 178 $\mu$m, 178 $\mu$m, and 159 $\mu$m. We must also perform a spatial average over the intensity profile of the chirped laser beam, a two-dimensional Gaussian with 1/e$^{2}$ radii of 144 $\mu$m and 116 $\mu$m. This averaging requires that the calculation be repeated for a large number of peak intensities,  I$_{0}$=$(1/2)(c\mu_{0})^{-1}|\epsilon_{0}|^{2}$, where c is the speed of light and $\mu_{0}$ is the permeability of free space. Finally, the results are summed over partial waves up to J=5: \begin{equation} R(I_{0}) = \sum^{5}_{J=0} (2J+1) R_{J}(I_{0}). \end{equation}

\section{3. Experiment}

\begin{figure}
    \centering 
    \includegraphics[width=8.25cm]{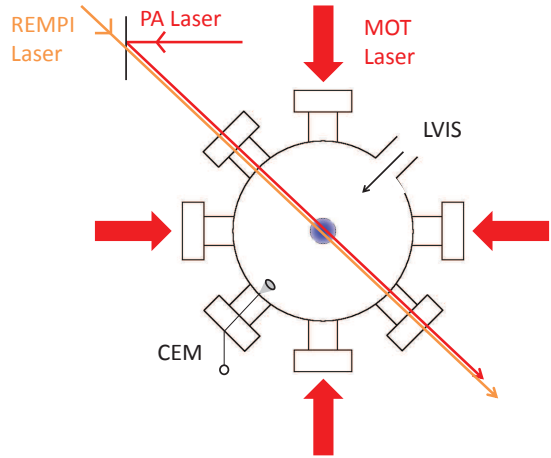} 
    \caption{Schematic of experimental set up. The cloud of ultracold atoms in the center of the vacuum chamber is produced by the MOT (only 4 of its 6 beams are shown) which is loaded by the low-velocity intense source (LVIS). Atoms are converted to molecules by the chirped photoassociation (PA) beam. The pulsed REMPI laser ionizes these molecules and the resulting ions are detected by the channel electron multiplier (CEM).}
\end{figure}

The experimental setup has been briefly described in our previous work \cite{Carini13,Carini15b} and is shown in Fig. 2. A magneto-optical trap (MOT) in the phase-stable configuration \cite{Rauschenbeutel98} is loaded with slow atoms from a low-velocity intense source (LVIS) \cite{Lu96}. The MOT provides a sample of N=2x10$^{6}$ ultracold $^{87}$Rb atoms at a temperature of $\sim$150 $\mu$K and peak density of 8x10$^{10}$ cm$^{-3}$. In the absence of photoassociation light, and at low density, the lifetime of atoms in the MOT is $\sim$70 s. The ultracold atoms are illuminated with pulses of frequency-chirped light in order to photoassociate (PA) them into ultracold $^{87}$Rb$_{2}$ molecules, as shown in Fig. 1. The frequency chirps are centered on the PA transition to 0$_{g}^{-}$ (v'=78), located 7.79 GHz below the $5S_{1/2} (F=2) \rightarrow 5P_{3/2} (F'=3)$ atomic transition \cite{Kemmann04,Carini13,Carini15b}. These excited molecules undergo either spontaneous emission (SPE) or stimulated emission (STE) and some fraction of them end up in high vibrational levels of the \textit{a} $^{3}\Sigma_{u}^{+}$ metastable state. Resonance-enhanced multiphoton ionization (REMPI) with a pulsed dye laser is used to detect these product molecules. Of particular interest is v''=39, the next-to-last vibrational level of  \textit{a} $^{3}\Sigma_{u}^{+}$. It is sufficiently weakly bound (764 MHz) that it can be populated by stimulated emission from the same chirp driving the PA step. 

\begin{figure}
    \centering 
    \includegraphics[width=8.25cm]{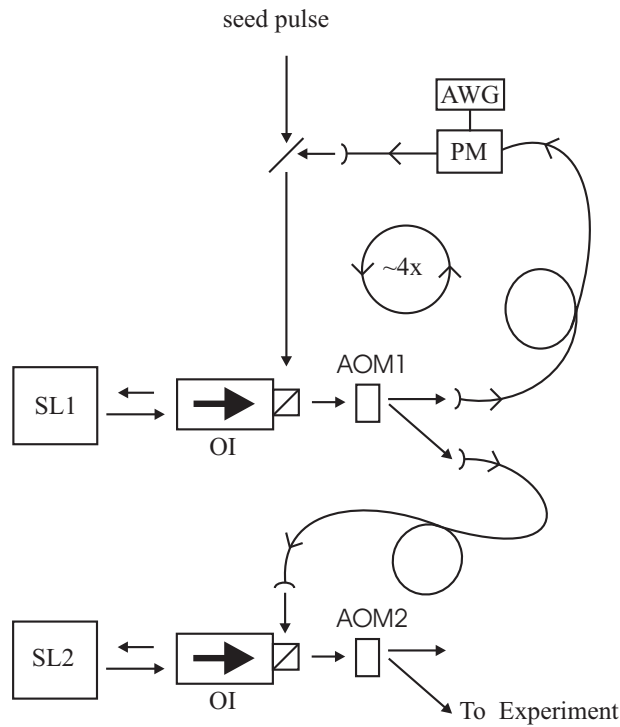} 
    \caption{Schematic of frequency chirp production. A pulse from the frequency-stabilized seed laser is used to injection lock slave laser 1 (SL1) by coupling through the output polarizing beamsplitter cube of an optical isolator (OI). The output of SL1 then passes through an acousto-optical modulator (AOM1) and is coupled into a fiber loop. This loop contains an electro-optical phase modulator (PM) driven by an arbitrary waveform generator (AWG). Light emerging from the fiber loop re-injection locks SL1 and is coupled back into the loop in order to accumulate the desired phase. After four traversals of the loop, AOM1 switches the light into a fiber whose output is used to injection lock SL2. AOM2 switches the output of SL2, thereby defining the pulse of chirped light sent to the experiment.}
\end{figure}

In order to produce fast frequency chirps, and to be able to shape these chirps, we use a high-speed fiber-based electro-optical phase modulator (EOSpace PM-0K1-00-PFA-PFA-790-S) driven by a 240 MHz arbitrary waveform generator (AWG), as shown in Fig. 3. Since frequency is the time derivative of phase, and the phase change produced by the modulator is proportional to the applied voltage, a programmed voltage waveform is chosen whose derivative yields the desired frequency vs. time. For example, a voltage which is quadratic in time, as shown as the PL curve in Fig. 4(a), yields a linear chirp. Because the maximum phase change achievable with the AWG/modulator combination is limited, we place the modulator in a fiber loop and traverse it multiple times to accumulate the desired phase shift \cite{Rogers07}. The starting optical frequency is set by a pulse from a 780 nm frequency-stabilized external-cavity diode seed laser which is used to injection lock a slave laser located in the fiber loop. The repetition rate of the AWG signal is synchronized to the $\sim$60 ns round trip time of the 7 m loop, and the slave laser is re-injection locked each time around the loop. After the desired phase is accumulated, requiring 4 passes, the light is switched out of the loop with an acousto-optical modulator (AOM). This chirped pulse then injection locks another high-power slave laser, which follows the chirp and whose output is switched on using a 200 MHz AOM to yield the final 15 ns FWHM pulse. When focused into the MOT, peak intensities up to $\sim$200 W/cm$^{2}$ are realized. We note that our time-domain chirp generation is quite different from the frequency-domain shaping techniques used for ultrafast pulses \cite{Weiner00}. In the former, we maintain the temporal duration and peak intensity of the pulse while increasing its bandwidth. In the latter, the bandwidth is not increased, but the pulse is stretched in time and its peak intensity reduced. 

\begin{figure}
    \centering 
    \includegraphics[width=8.25cm]{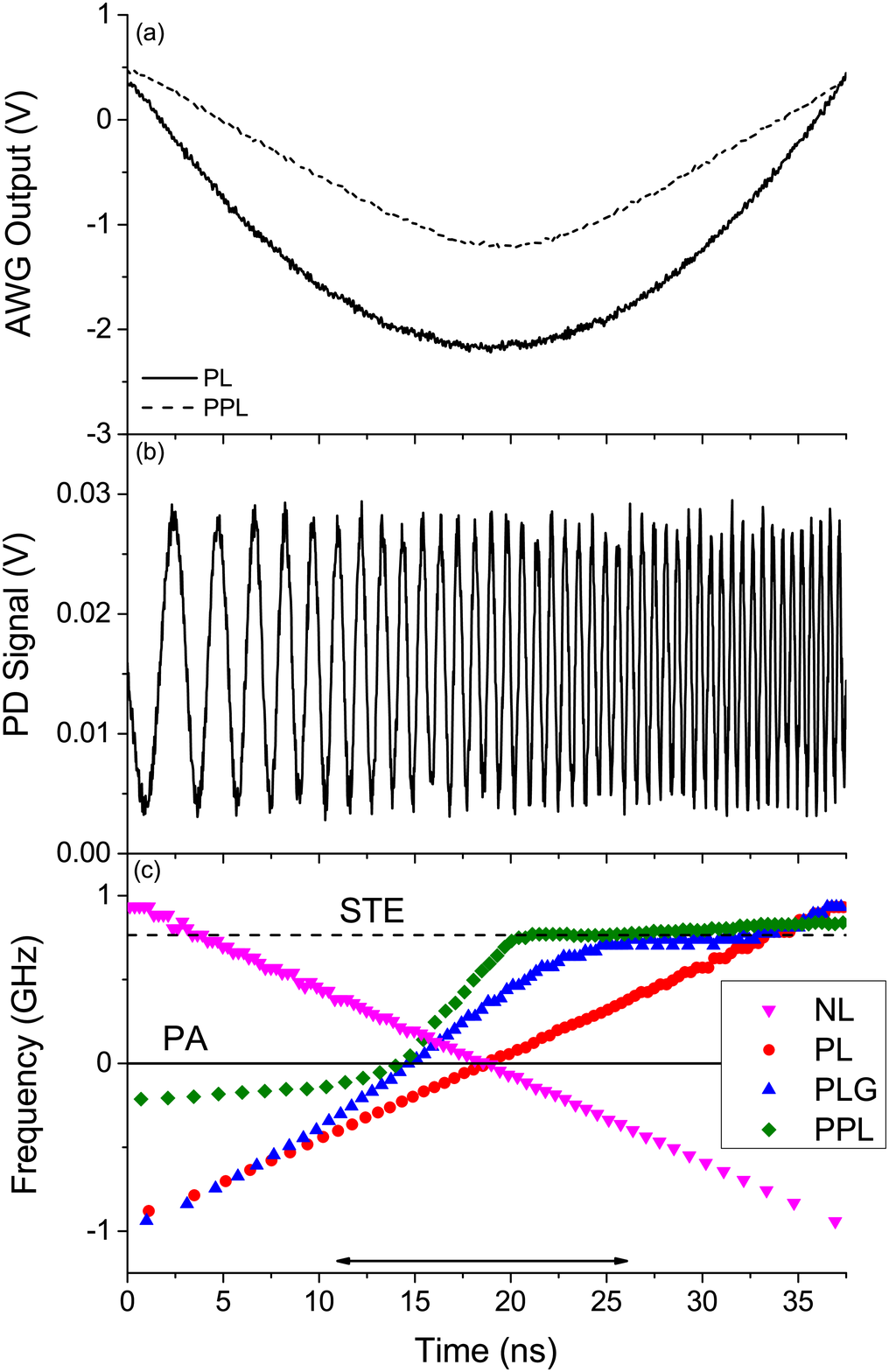} 
    \caption{(a) Voltage vs. time applied to phase modulator in order to produce the positive linear chirp (PL) and positive piecewise linear chirp (PPL). (b) Heterodyne signal for the PL chirp. (c) Frequency vs. time for the various chirps used: negative linear (NL); positive linear (PL); positive linear plus Gaussian (PLG); and positive piecewise linear (PPL). The solid and dashed horizontal lines represent the PA transition to 0$_{g}^{-}$ (v'=78) and the STE transition to \textit{a} $^{3}\Sigma_{u}^{+}$ (v''=39), respectively. For the unchirped (UC) case, the light remains on resonance with the PA transition. The horizontal double-ended arrow denotes the 15 ns FWHM Gaussian intensity pulse centered at t=18.75 ns.}
\end{figure}

The frequency chirps are measured by combining the chirped light with a fixed frequency reference laser and measuring the resulting heterodyne signal (Fig. 4(b)) with a 2 GHz photodiode (Thorlabs SV2-FC) and a 8 GHz oscilloscope. The various chirp shapes used in the present work are shown in Fig. 4(c): linear chirps, both positive linear (PL) and negative linear (NL), with slopes of $\pm$1.9 GHz in 37.5 ns; a positive linear chirp with a Gaussian (0.425 GHz amplitude, 15 ns FWHM) superimposed (PLG); and a positive piecewise linear (PPL) chirp comprising gently sloping ($\sim$10 MHz/ns) initial and final segments with a steep ($\sim$120 MHz/ns) rise in between. As shown by the double-ended arrow in Fig. 4(c), the central portions of these chirps are selected, using the final AOM, with a 15 ns FWHM Gaussian intensity pulse. For the linear chirps, the time scales for both the chirps and pulse widths have been reduced by a factor of $\sim$2.7 from our earlier work \cite{Carini13}, thereby minimizing the role of spontaneous emission. Also the chirp range has been widened by a factor of $\sim$2, allowing both the absorption (PA) and stimulated emission (STE) steps to occur within the high intensity portion of the pulse.

The REMPI detection of the ultracold molecules employs a pulsed dye laser pumped by the second harmonic of a pulsed Nd:YAG laser operating at 10 Hz. The wavelength is tuned to a broad feature in the REMPI spectrum at $\sim$16608.5 cm$^{-1}$. Based on previous work \cite{Kemmann04,Gabbanini00}, this wavelength effectively ionizes high vibrational levels of the \textit{a} $^{3}\Sigma_{u}^{+}$ state. The $\sim$0.2 cm$^{-1}$ laser bandwidth, measured using two-photon excitation to atomic Rydberg states, does not allow the high-v'' levels of \textit{a} $^{3}\Sigma_{u}^{+}$ to be resolved. Pulses of light, 5 ns in duration and $\sim$4.8 mJ in energy, are focused onto the cold atom cloud. The Gaussian atomic cloud has dimensions (average 1/e$^{2}$ radius) of 172 $\mu$m, while the REMPI beam is larger, $\sim$3 mm in diameter, to provide better overlap with the untrapped and ballistically expanding cloud of ultracold molecules. Ions produced by REMPI are accelerated into a Channeltron detector. A digital boxcar averager enables the Rb$_{2}^{+}$ ions of interest to be separated from background Rb$^{+}$ by their time of flight (2.3 $\mu$s vs. 1.7 $\mu$s) to the detector. The MOT light is extinguished for 50 $\mu$s centered on the REMPI pulse in order to minimize this Rb$^{+}$ background.

	The number of molecules produced in a single chirped pulse is quite small, so these pulses are repeated every 450 ns in order to build up the signal for each REMPI pulse. The time-averaged molecule formation rate R, which we can compare to the simulations, is the product of the number of molecules produced per chirped pulse and the pulse repetition rate. Since we are using a sequence of chirped pulses, we must account for the photodestruction of already existing molecules by subsequent pulses, which occurs at a time-averaged rate $\Gamma_{PD}$ per molecule. In addition, since they are not trapped by the MOT, the molecules escape ballistically from the detection region at a rate $\Gamma_{esc}$ = 100(4) s$^{-1}$. This rate is measured using molecules produced by MOT light (i.e., with no chirped pulses present), as shown in the inset to Fig. 5. We determine R by measuring the REMPI signal as the formation time (number of chirped pulses) is varied.  The number N of detectable molecules per REMPI pulse, which increases at a constant rate R and decreases at a rate -($\Gamma_{PD}$ + $\Gamma_{esc}$)N,  evolves according to:
\begin{equation}
N(t) = \frac{R}{\Gamma_{PD}+\Gamma_{esc}}(1 - e^{-(\Gamma_{PD}+ \Gamma_{esc})t}).	\label{evolve}
\end{equation}
By fitting the exponential approach to steady state, as shown in Fig. 5, we determine the total loss rate, $\Gamma_{PD}$ + $\Gamma_{esc}$. Combining this with the measured steady-state number of detectable molecules per REMPI pulse, N$_{SS}$ = R/($\Gamma_{PD}$ + $\Gamma_{esc}$), allows us to extract R, the quantity of interest. Comparing the two intensities in Fig. 5, we see that for the higher intensity, the approach to steady state is faster, indicating a larger value of $\Gamma_{PD}$, and the steady-state value is also larger, indicating a higher formation rate R.

\begin{figure}
    \centering 
    \includegraphics[width=8.25cm]{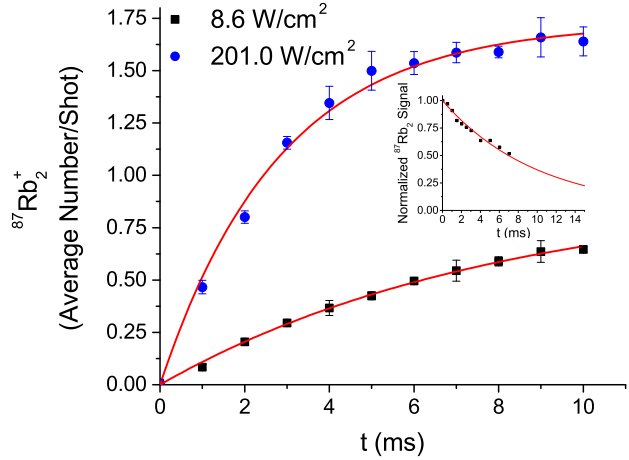} 
    \caption{Molecular ion signal vs. illumination time (number of chirped pulses) for the positive linear chirp for two peak intensities: 8.6 W/cm$^{2}$ and 201.0 W/cm$^{2}$. Solid lines are fits to Eq. \ref{evolve}. Inset: decay of MOT-formed molecule signal due to ballistic escape. The solid line is a fit to an exponential, yielding a decay rate of $\Gamma_{esc}$ = 0.100 $\pm$ 0.004 ms$^{-1}$.}
\end{figure}

 In Fig. 6, we plot the molecular formation rate R, extracted from fits to Eq. \ref{evolve} as shown in Fig. 5, vs. the peak intensity. The various curves represent the different chirp shapes shown in Fig. 4(c): unchirped (UC); negative linear (NL); positive linear (PL); positive linear with a Gaussian superimposed (PLG); and positive piecewise linear (PPL). There is an obvious dependence of formation rate on chirp shape. All of the positive chirps outperform the negative chirp. This was seen even in our earlier work \cite{Carini13} on slower time scales, and attributed to stimulated emission from the excited state into \textit{a} $^{3}\Sigma_{u}^{+}$. In contrast to our earlier work, the positive linear chirp does as well as, or perhaps even outperforms, the unchirped pulse. Although it incorporates some degree of shaping, the positive linear plus Gaussian chirp does only about as well as the positive linear chirp. The most dramatic feature of Fig. 6 is the improvement provided by the positive piecewise linear chirp. This demonstrates the benefit which can be provided by judicious shaping of the frequency chirp.

\begin{figure}
    \centering 
    \includegraphics[width=8.25cm]{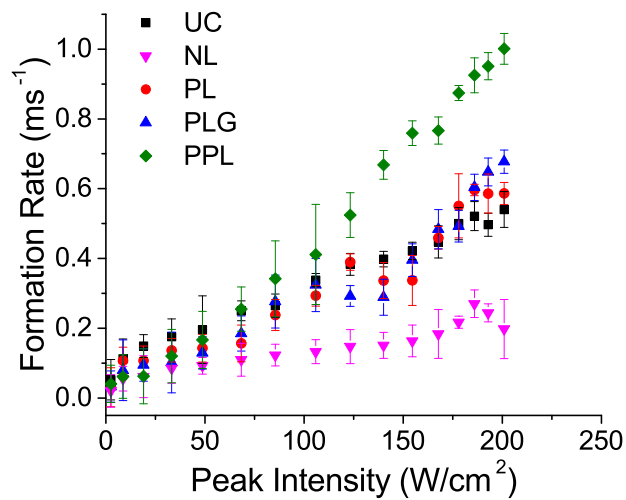} 
    \caption{Measured molecular formation rates vs. peak intensity for the various chirp shapes shown in Fig. 4(c).}
\end{figure}

\section{4. Comparison of Theory and Experiment}

\begin{figure}
    \centering 
    \includegraphics[width=8.25cm]{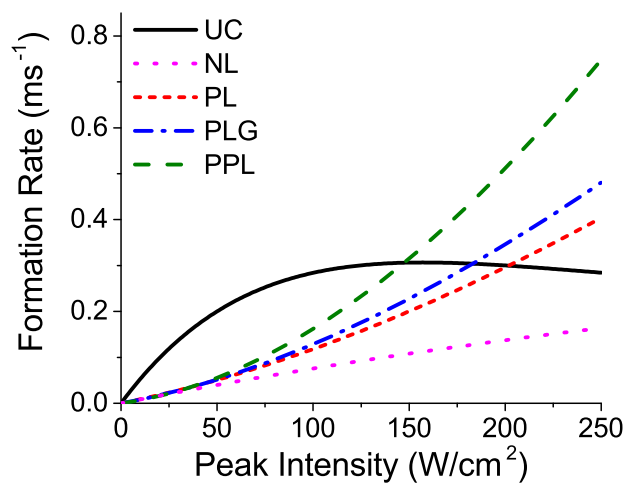} 
    \caption{Simulated molecular formation rates vs. peak intensity for the various chirp shapes shown in Fig. 4(c).}
\end{figure}

The molecular formation rates for the various chirp shapes in Fig. 4(c), smoothed with a 2 ns FWHM Gaussian, are calculated according to the procedure outlined in Sect. 2.  The results are plotted as functions of intensity in Fig. 7. With the exception of the unchirped pulses, there is generally good agreement between these results of the simulations and the experimental data shown in Fig. 6. In comparing the absolute rates, there is significant systematic uncertainly in the simulated rates due to factor of $\sim$2 uncertainties in both the trapped atom number and atomic density calibrations. The relative efficiencies of the various chirp shapes are consistent between theory and experiment: NL is the least efficient, PPL the most efficient, and PL and PLG occupy the middle ground.

In order to understand the relative efficiencies of the various chirps, we examine the calculated time evolution of the populations of the relevant states for a fixed intensity and collision energy, and a single partial wave (J=0). We focus on the 0$_{g}^{-}$ (v'=78) level which is populated by PA, and the \textit{a} $^{3}\Sigma_{u}^{+}$ (v''=39) target state. For the latter, we separate out the contributions of stimulated emission (STE) and spontaneous emission (SPE). As mentioned in Sect. 2, the SPE contribution includes other v'' levels and is therefore somewhat of an overestimate.  Fig. 8 shows these three populations as functions of time for the PL and PPL chirps at two different intensities. We note that the other vibrational levels of 0$_{g}^{-}$ in our basis set, as well as the levels of 1$_{g}$, are all included in the calculation. Their populations are not shown because their contributions to detectable molecules are minimal. For example, compared to the 0$_{g}^{-}$ SPE contributions shown in Fig. 8(d), the 1$_{g}$ contributions are smaller by a factor of $\sim$10$^{6}$, due mainly to the much smaller FCFs to \textit{a} $^{3}\Sigma_{u}^{+}$ (v''=39).

	For both the PL and PPL chirp shapes, we also show (Figs. 8(a) and 8(e)) the energy levels in the dressed picture. In this picture, the excited energies are constant (horizontal lines), while the time-dependent (chirped) photon energy is added to energies of the \textit{a} $^{3}\Sigma_{u}^{+}$ (v''=39) continuum and \textit{a} $^{3}\Sigma_{u}^{+}$, yielding the two parallel curves offset vertically by the v''=39 binding energy. The light is resonant with a transition between two states when their curves cross. For example, in Fig. 8(a), points B and C indicate the PA and STE resonances, respectively. These crossings are actually avoided crossings due to the laser coupling of the states. However, the resulting gaps are too small to be seen on this scale and are therefore not shown.

\begin{figure}
    \centering 
    \includegraphics[width=16.5cm]{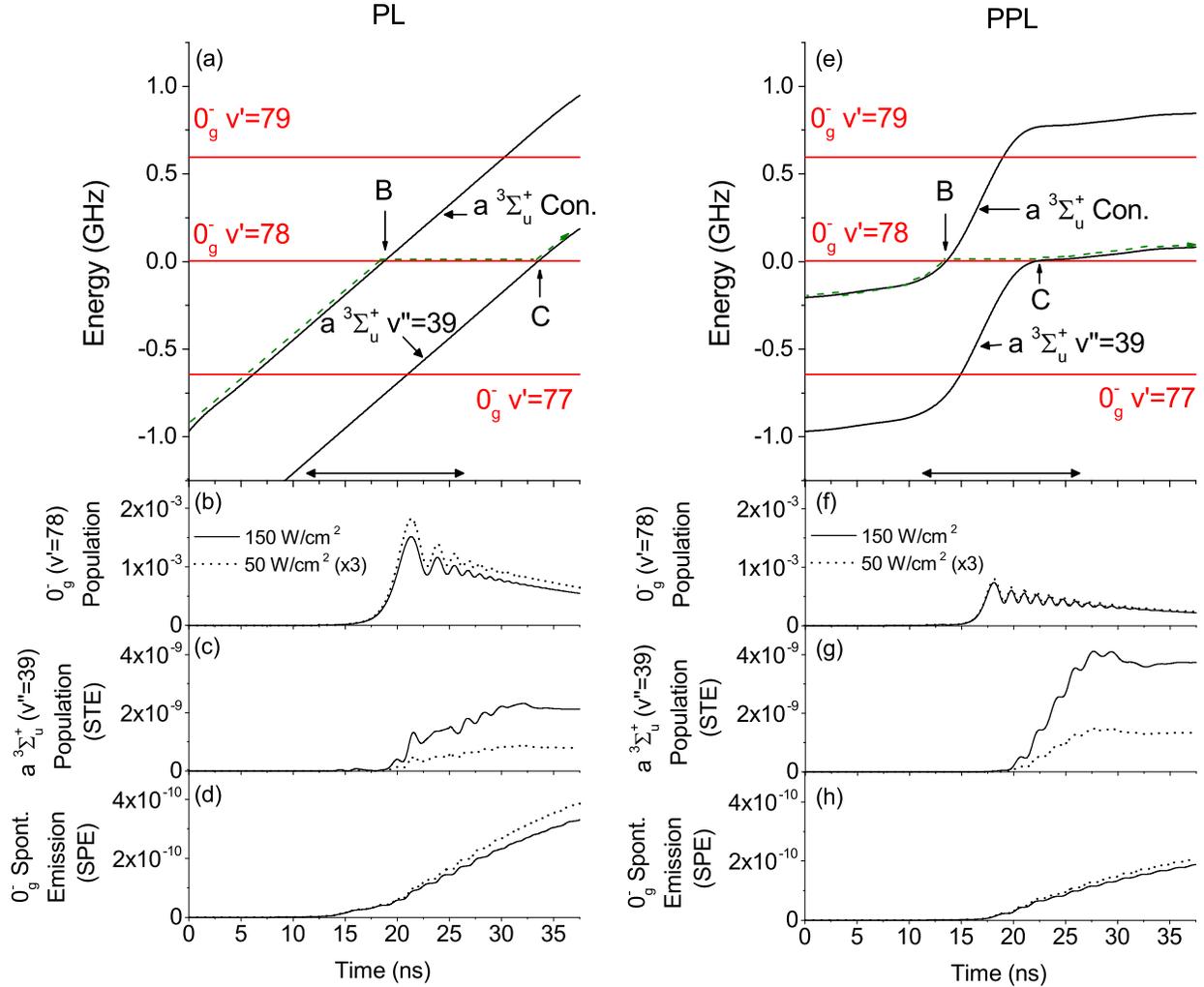} 
    \caption{The dressed state molecular energies are shown in (a) for the positive linear (PL) chirp. Here the horizontal lines indicate the various vibrational levels of the 0$_{g}^{-}$ excited state, while the solid sloped lines indicate the energy of the continuum (upper curve) and the v''=39 bound level (lower curve) of the \textit{a} $^{3}\Sigma_{u}^{+}$ state, with the energy of the chirped photon added. The constant energy offset between these curves is the v''=39 binding energy. The dashed line shows the adiabatic path from the free-atom continuum to \textit{a} $^{3}\Sigma_{u}^{+}$ (v''=39), via 0$_{g}^{-}$ (v'=78). The curve crossings indicated by B and C are the PA and STE resonances, respectively. The horizontal double-ended arrow denotes the 15 ns FWHM Gaussian intensity pulse centered at t=18.75 ns. Simulated level populations vs. time for the positive linear (PL) chirp at peak intensities of 50 W/cm$^{2}$ (dotted curves) and 150 W/cm$^{2}$ (solid curves): (b) v'=78 excited state; (c) v''=39 resulting from stimulated emission (STE); (d) v''=39 resulting from spontaneous emission (SPE). The curves for 50 W/cm$^{2}$ are multiplied by a factor of 3.  (e)-(h) Same as (a)-(d) for the positive piecewise linear (PPL) chirp.}
\end{figure}

	Examining the populations in Fig. 8, we see the following. For both chirps, the 0$_{g}^{-}$ (v'=78) excited-state population (Figs. 8(b) and 8(f)) begins to appear shortly after the PA resonance. This population is less for the PPL chirp because it passes through the PA resonance earlier in time, and therefore at a lower intensity. As the chirps proceed, the transfer to \textit{a} $^{3}\Sigma_{u}^{+}$ (v''=39) by STE begins to occur (Figs. 8(c) and 8(g)). Here the PPL chirp does much better. It reaches the STE resonance earlier, and therefore at a higher intensity, and also traverses it with a smaller slope. This combination makes the transfer more adiabatic. The final STE population for the PPL chirp exceeds that for the PL chirp by a factor of $\sim$2, despite the fact that its excited-state population was significantly smaller. At our intensities, STE has a negligible effect on the excited-state population, so the SPE populations (Fig. 8(d) and 8(h)) simply follow exponential decay of the excited states. At long times, allowing for complete decay, the PL and PPL SPE populations for 150 W/cm$^{2}$ reach 6x10$^{-10}$ and 3x10$^{-10}$, respectively. For the PPL chirp this SPE contribution is at least an order of magnitude less than that of STE, showing that the formation in this case is predominantly coherent.
	
	Comparing the various populations for the two peak intensities, 50 W/cm$^{2}$ and 150 W/cm$^{2}$, we see that for both chirp shapes, the excited-state populations and SPE contributions scale with intensity, while the STE scales much faster, consistent with it being a two-photon process.
	
	The general shape for the PPL chirp was inspired by earlier calculations we performed using local control of the phase to optimize the molecule formation \cite{Carini15}.  Unlike many coherent control schemes, local control \cite{Marquetand07,Engel09} is not iterative, but unidirectional, adjusting the field (its phase in our case) at each time step in order to optimize the target (molecule formation in our case) at the next step. There are two advantages to local control for our application: 1) since the long time scales make our computations rather intensive, multiple iterations, for example in a genetic algorithm, would be very expensive; and 2) the resulting optimizing waveforms can be simple, and therefore experimentally feasible. A simple waveform was indeed the outcome for our situation. The optimal frequency evolution was a rapid jump, halfway through the pulse, from the PA resonance to the STE resonance. Because of speed limitations in our chirp production, and uncertainties in locating and maintaining the resonances, we used the PPL chirp as a stand-in for the step-function chirp, with positive results.   
			
	The one unsatisfying aspect of the comparison between our simulations and experimental data is the poor match for the unchirped (UC) pulses, especially for low intensities. Although we have no good explanation for this behavior, the tuning of the laser to the PA resonance, and possible small-scale structure in this resonance, are more critical when there is no chirp.

\section{5. Prospects for Further Improvements}

	Examining Fig. 7, it is obvious that the molecule formation rate is a strong function of intensity. In fact, the dependence for the positive chirps is close to quadratic, as expected for a two-photon process: PA followed by STE. We extend the calculations to much higher peak intensities, up to 2x10$^{4}$ W/cm$^{2}$, in Fig. 9. For the positive chirps (PL, PLG, PPL), the formation rates at low intensities (<30 W/cm$^{2}$) are linear with intensity, consistent with a one-photon process: PA followed by SPE. This is the regime where SPE dominates over STE. At higher intensities, STE becomes dominant and the intensity dependence is close to quadratic, until eventually saturating. If we could increase the intensity by a factor of 10 from our current maximum of 201 W/cm$^{2}$, we would gain a factor of 100 in formation rate for the PPL chirp. For the unchirped (UC) and negative linear (NL) chirp cases, the intensity dependence is linear until saturating. This is consistent with STE playing a minimal role. We note that for these calculations of formation rates, we are assuming a constant density profile of the MOT with negligible depletion.

\begin{figure}
    \centering 
    \includegraphics[width=8.25cm]{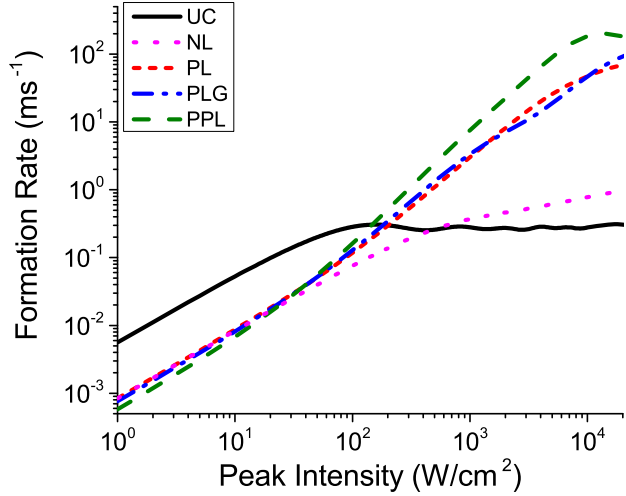} 
    \caption{Molecular formation rate for the various chirp shapes vs. peak intensity up to 2x10$^{4}$ W/cm$^{2}$. Note that both the horizontal and vertical axes are logarithmic. Parameters are the same as in Fig. 7.}
\end{figure}

The time-dependent populations for the PPL chirp at high intensity (1x10$^{4}$ W/cm$^{2}$) are shown in Fig. 10. Comparing to the case of much lower intensity (150 W/cm$^{2}$) shown in Fig. 8, we see that STE increasingly dominates over SPE, meaning that the process becomes almost completely coherent at high intensities. This is shown explicitly in Fig. 10(d), where we plot the ratio of populations from STE and SPE.

\begin{figure}
    \centering 
    \includegraphics[width=8.25cm]{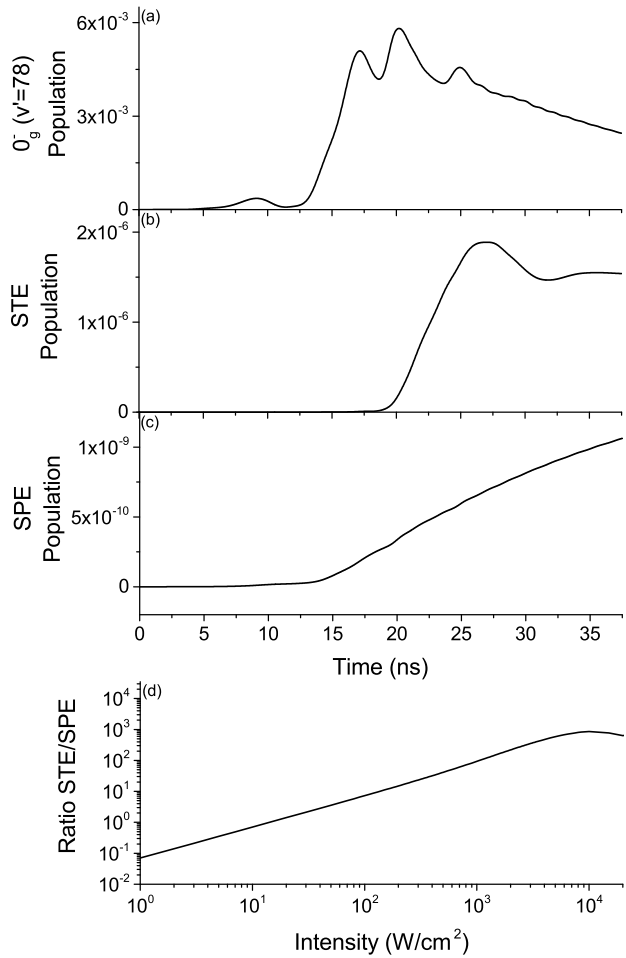} 
    \caption{(a)-(c) Populations vs. time for the PPL chirp, as in Fig. 8(e)-(g), but for a peak intensity of 1x10$^{4}$ W/cm$^{2}$. The ratio of STE to SPE populations vs. peak intensity is shown in (d). Note that in (d), both the horizontal and vertical axes are logarithmic.}
\end{figure}

We have also used the simulations to explore how the choice of intermediate (excited) state affects the formation rate. The intermediate state used in recent experiments \cite{Carini13,Carini15b}, 0$_{g}^{-}$ (v'=78), was chosen because the PA step is close to the asymptote and therefore quite strong. This leads to readily observable trap loss for quasi-cw unchirped light, allowing the PA resonance to be easily located. However, the second step in coherent molecule formation, STE down to \textit{a} $^{3}\Sigma_{u}^{+}$ (v''=39), is rather weak for this intermediate state. For example, if we compare Figs. 10(a) and 10(b), we see that <10$^{-3}$ of the excited-state population is transferred by STE, even at this high intensity of 10$^{4}$ W/cm$^{2}$. Also, the PA step begins to saturate at high intensities. In order to optimize the two-step coherent formation of \textit{a} $^{3}\Sigma_{u}^{+}$ (v''=39), we need to maximize the product of the strengths of the two transitions, PA and STE. This strategy is similar to that for optimizing stimulated Raman transfer of ultracold molecules from high vibrational levels to the vibrational ground state \cite{Kim11}. Towards this end, we plot in Fig. 11 the Franck-Condon factors (FCFs) for the PA step from the continuum (s-wave scattering state at 150 $\mu$K) to 0$_{g}^{-}$ (v'), and for the STE step from 0$_{g}^{-}$ (v') to \textit{a} $^{3}\Sigma_{u}^{+}$ (v''=39), as functions of v'. The product of these FCFs is also shown. It is seen that for v'=78, which is located on a broad local maximum, the PA step is strong, but the STE step is very weak, leading to poor overall efficiency. A much better choice is v'=31, which despite yielding a much lower FCF$_{PA}$, improves the product of FCFs by over two orders of magnitude.

\begin{figure}
    \centering 
    \includegraphics[width=8.25cm]{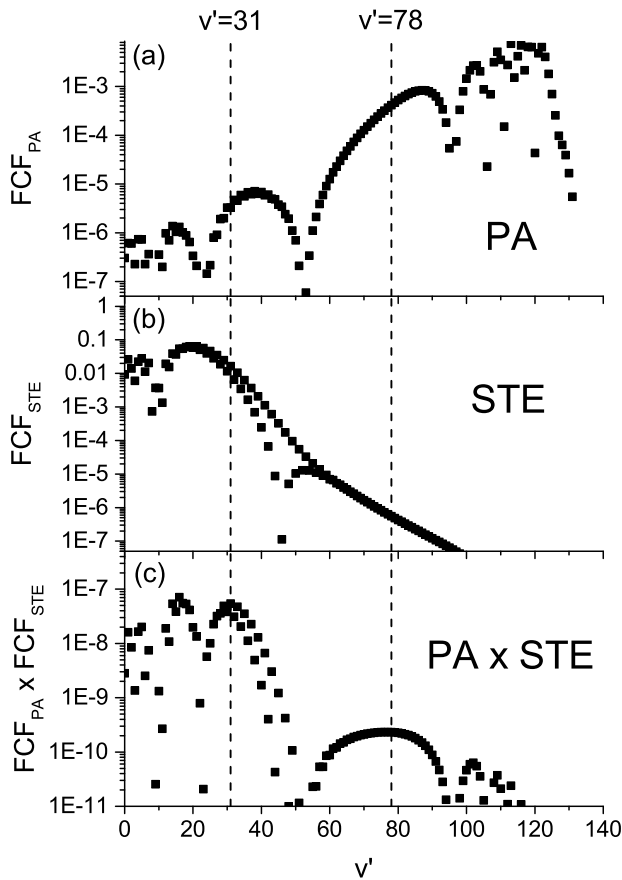} 
    \caption{Franck-Condon factors (FCFs) vs. v' for: (a) the photoassociation (PA) step from the continuum to 0$_{g}^{-}$ (v'); (b) the stimulated emission (STE) step from 0$_{g}^{-}$ (v') to \textit{a} $^{3}\Sigma_{u}^{+}$ (v''=39). (c) The product of the FCFs shown in (a) and (b). Note that the vertical axes are logarithmic.}
\end{figure}

This tradeoff of the two FCFs can be understood by examining the wavefunctions of the states involved, as shown in Fig. 12. We see that the long-range 0$_{g}^{-}$ (v'=78) excited state (Fig. 12(b)) has very good overlap with the initial continuum state (Fig. 12(a)), but poor overlap with the relatively short-range  \textit{a} $^{3}\Sigma_{u}^{+}$ (v''=39) target state (Fig. 12(d)). On the other hand,  0$_{g}^{-}$ (v'=31) (Fig. 12(c)) has reasonable overlap with both.          

\begin{figure}
    \centering 
    \includegraphics[width=8.25cm]{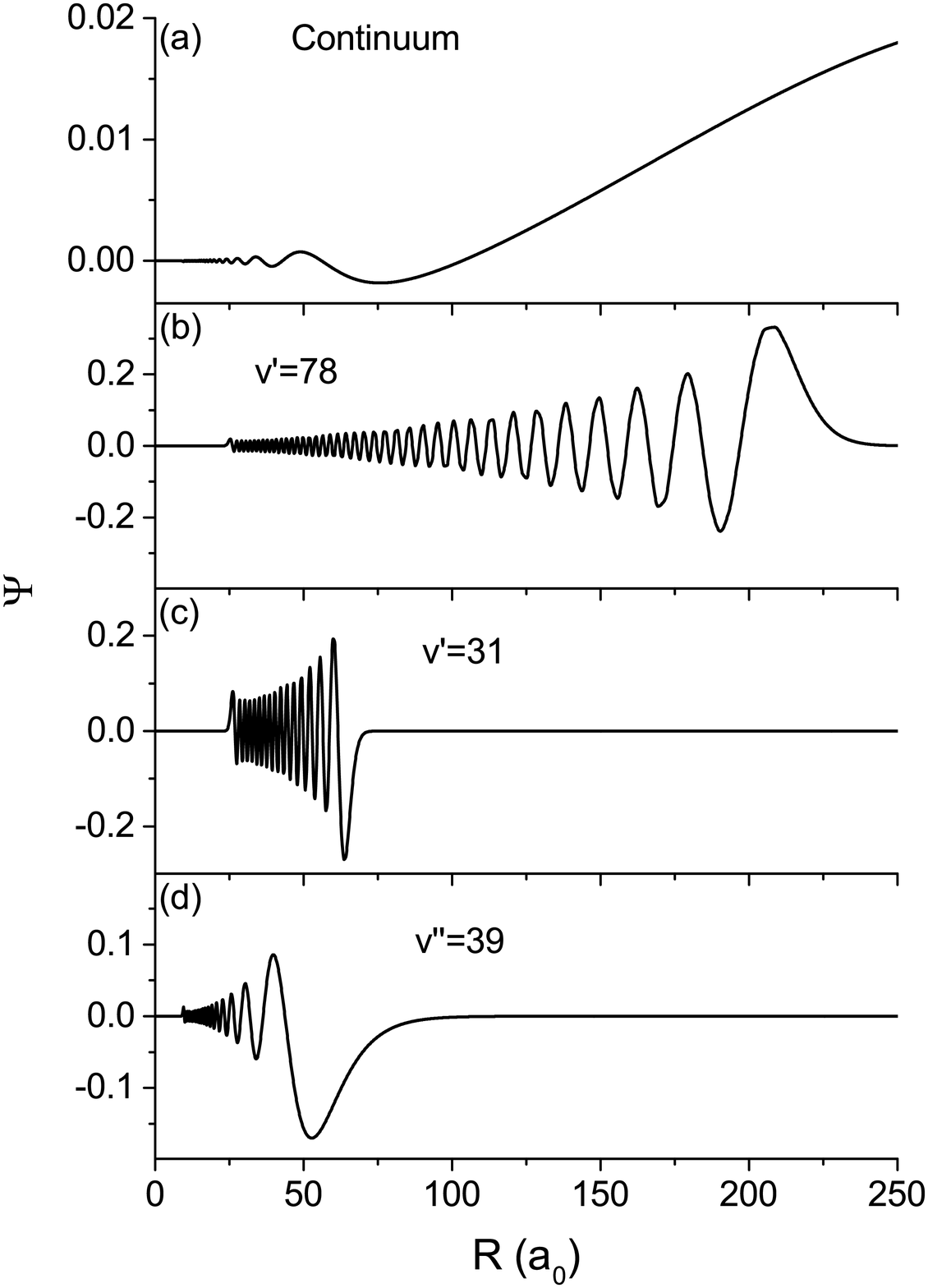} 
    \caption{Wavefunctions for: (a) the \textit{a} $^{3}\Sigma_{u}^{+}$ s-wave continuum at 150 $\mu$K; (b) 0$_{g}^{-}$ (v'=78); (c) 0$_{g}^{-}$ (v'=31); (d) \textit{a} $^{3}\Sigma_{u}^{+}$ (v''=39). Note that (a) overlaps well with (b), while (d) overlaps well with (c).}
\end{figure}

	The potential gain in molecular formation rate offered by optimizing the intermediate state is shown in Fig. 13. This is the same plot as Fig. 9, but with 0$_{g}^{-}$ (v'=31) instead of 0$_{g}^{-}$ (v'=78) as the intermediate state. For these calculations, the excited state basis is expanded to $\sim$115 GHz which includes 12 levels each of 0$_{g}^{-}$ and 1$_{g}$. We see that for the positive chirps, we gain at least two orders of magnitude. For example, at 150 W/cm$^{2}$, the gain for the PPL chirp is a factor of 298. Interestingly, the relative efficiencies of the different positive chirps shapes are modified. Here, the PLG chirp slightly outperforms the PPL chirp. Also of note is the fact that the rates for the positive chirps are quadratic with intensity down to very low intensities, indicating the almost completely coherent nature of the process. This is verified by examining, in Fig. 14, the various populations when v'=31 is used as the intermediate state for the PPL chirp. This can be directly compared to the solid curves in Figs. 8(f)-8(h), where v'=78 was used as the intermediate state. We see that both the peak intermediate (excited) state population and the SPE contribution to \textit{a} $^{3}\Sigma_{u}^{+}$ are reduced by a factor of 10 when using v'=31, while the STE contribution to \textit{a} $^{3}\Sigma_{u}^{+}$ (v''=39) is increased by a factor of 222. In Fig. 14(d), we plot the ratio of the contributions of STE and SPE to the target state when 0$_{g}^{-}$ (v'=31) is used as the intermediate state. STE dominates over the entire intensity range, confirming that the process is coherent. Comparing to Fig. 10(d), which shows the same STE/SPE ratio but for the 0$_{g}^{-}$ (v'=78) intermediate state, we see how much more coherent the process is for v'=31. As discussed in Sect. 2, for the 0$_{g}^{-}$ (v'=31) intermediate state, a significant fraction of the SPE population may be undetected.  Therefore, the v'=31 SPE populations used in these plots may be an overestimate.

\begin{figure}
    \centering 
    \includegraphics[width=8.25cm]{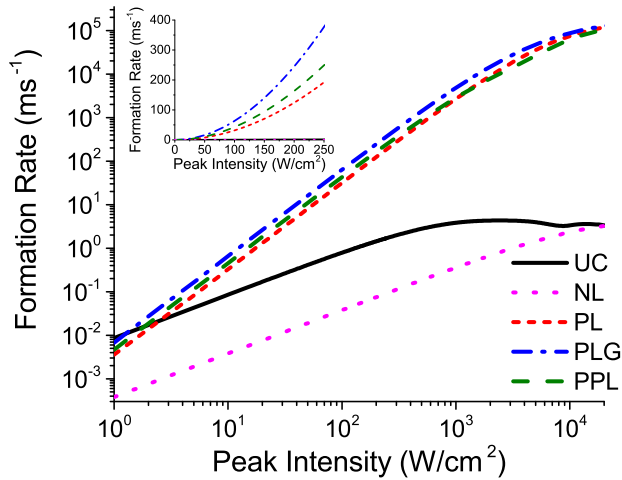} 
    \caption{Molecular formation rate vs. peak intensity up to 2x10$^{4}$ W/cm$^{2}$ using  0$_{g}^{-}$ (v'=31) as the intermediate state. Note that both the horizontal and vertical scales are logarithmic. The inset is the same plot for lower intensities with linear scales.}
\end{figure}

\begin{figure}
    \centering 
    \includegraphics[width=8.25cm]{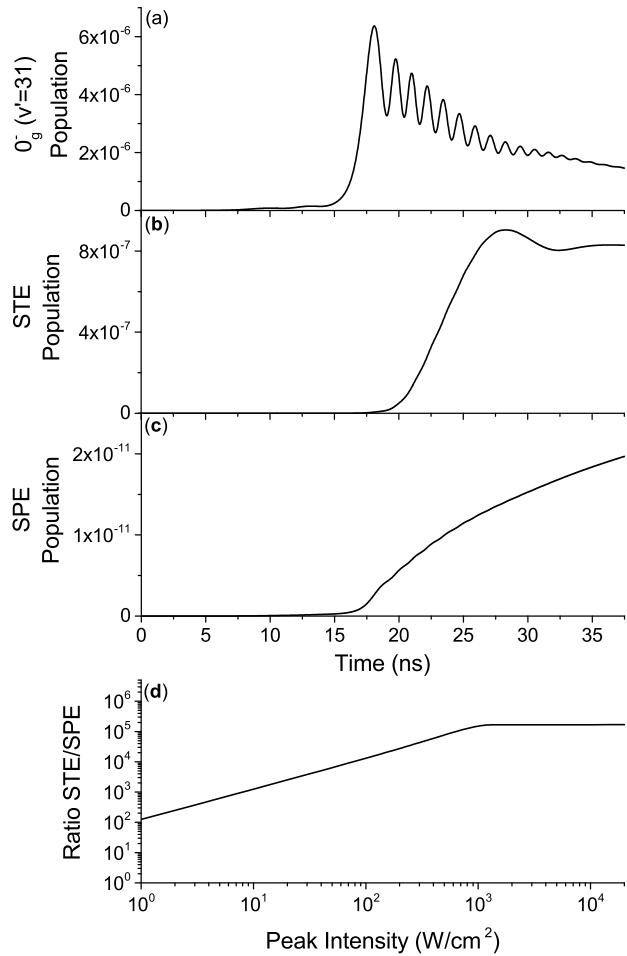} 
    \caption{Populations vs. time for the PPL chirp for a peak intensity of 150 W/cm$^{2}$, as in Figs. 8(e)-(g), but using 0$_{g}^{-}$ (v'=31) as the intermediate state. The ratio of STE to SPE populations vs. peak intensity is shown in (d). Note that in (d), both the horizontal and vertical axes are logarithmic.}
\end{figure}

	Although both higher intensity and a more favorable intermediate state are predicted to lead to improved molecule formation rates, they may also result in a higher rate of photodestruction of already existing molecules. Since photodestruction should be only linear in intensity, we expect a net gain in the number of detectable molecules as we go to higher intensities. This is a topic which warrants further investigation.

\section{6. Summary}

We have described the use of shaped frequency chirps to form ultracold molecules. The first step is the production of excited molecules by photoassociation (PA). Molecules in a high vibrational level of the \textit{a} $^{3}\Sigma_{u}^{+}$ are then formed by either spontaneous emission (SPE) of the excited state or stimulated emission (STE) by the chirped light. Proper time ordering of the PA and STE processes dictates that a positive chirp should be used. The shape of the chirp can have a dramatic influence on the relative importance of STE and SPE processes, as well as the overall formation rate. We have provided details on recent experiments, as well as the corresponding quantum simulations, and used the time-dependent state populations to confirm the physical picture of coherent molecule formation. Paths for further improvements in efficiency have been identified: higher intensities; and different intermediate states.

\begin{acknowledgement}

	This work is supported by the U.S. Department of Energy Office of Science, Office of Basic Energy Sciences, Chemical Sciences, Geosciences, and Biosciences Division under Award Number DE-FG02-92ER14263. We also acknowledge support from the US-Israel Binational Science Foundation through grant number 2012021. 

\end{acknowledgement}

%
%

\bibliography{library}

\end{document}